\documentclass[twocolumn,showpacs,preprintnumbers,amsmath,amssymb]{revtex4}

\usepackage{graphicx}% Include figure files
\usepackage{dcolumn}
\usepackage{bm}% bold math
\usepackage{epsfig}
\begin{document}
\title{Optical properties of SiC nanotubes: A systematic $\textit{ab initio}$ study}
\author{I. J. Wu and G. Y. Guo\footnote{Electronic address: gyguo@phys.ntu.edu.tw}}
%\author{I. J. Wu and G.\ Y.\ Guo\footnote{Electronic address: gyguo@phys.ntu.edu.tw}}
\affiliation{
Department of Physics and Center for Theoretical Sciences, National Taiwan University, Taipei, Taiwan 106, R.O.C.}

\date{\today}

\begin{abstract}
%A systematic $\textit{ab initio}$ study of the electronic and optical properties of 
%SiC nanotubes (NTs) has been performed within density functional theory with the local 
%density approximation. In particular, 
The band structure and optical dielectric function $\epsilon$ of single-walled zigzag 
$[$(3,0),(4,0),(5,0),(6,0),(8,0),(9,0),(12,0),(16,0),(20,0),(24,0)$]$, 
armchair $[$(3,3),(4,4),(5,5),(8,8),(12,12),(15,15)$]$,
and chiral $[$(4,2),(6,2),(8,4),(10,4)$]$ SiC-NTs as well as the single honeycomb SiC 
sheet have been calculated within density functional theory with the local
density approximation. The underlying atomic structure of the SiC-NTs is 
determined theoretically. 
It is found that all the SiC nanotubes are semiconductors, except
the ultrasmall (3,0) and (4,0) zigzag tubes which are metallic.
%may be regarded as the thinnest conducting SiC nanowires. 
Furthermore, the band gap
of the zigzag SiC-NTs which is direct, may be reduced from that of the SiC sheet
to zero by reducing the diameter ($D$), though the band gap
for all the SiC nanotubes with a diameter larger than
$\sim$20 \AA$ $ is almost independent of diameter. 
%Furthermore, all the semiconducting zigzag SiC-NTs have a direct band gap. 
%The optical properties of the SiC-NTs
%can be divided into two spectral regions, namely,
%the low-energy region (0$\sim$6 eV), in which the optical transitions involve mainly
%the $\pi$-bands, and the high-energy region from 6 eV upwards, where interband
%transitions involve mainly the $\sigma$-bands.
For the electric field parallel to the tube axis ($E\parallel \hat{z}$), 
the $\epsilon''$ for all the SiC-NTs
with a moderate diameter (say, $D$ $>$ 8 \AA$ $)
% for the zigzag SiC-NTs, $D$ $>$ 6 \AA$ $ for the chiral SiC-NTs) 
in the low-energy region (0$\sim$6 eV) consists of a single
distinct peak at $\sim$3 eV.
However, for the small diameter SiC nanotubes such as the (4,2),(4,4) SiC-NTs,
the $\epsilon''$ spectrum does deviate markedly
from this general behavior.  In the high-energy region (from 6 eV upwards), the $\epsilon''$
for all the SiC-NTs exhibit a broad peak centered at $\sim$7 eV.
For the electric field perpendicular to the tube axis ($E\perp \hat{z}$), 
the $\epsilon''$ spectrum of all the SiC-NTs except
the (4,4), (3,0) and (4,0) nanotubes, in
the low energy region also consists of a pronounced peak at around 3 eV
whilst in the high-energy region is roughly made up of a broad hump starting
from 6 eV. The magnitude of the peaks is in general about
half of the magnitude of the corresponding ones for $E\parallel \hat{z}$. 
%showing a moderate optical anisotropy in the SiC-NTs.
Interestingly, the calculated static dielectric
constant $\epsilon(0)$ for all the SiC nanotubes
is nearly independent of diameter and chirality
with $\epsilon(0)$ for $E\parallel \hat{z}$ being only about 30 \% larger
than for $E\perp \hat{z}$.
The calculated electron energy loss spectra of all the SiC nanotubes
for both electric field
polarizations are rather similar to that of $E\perp \hat{c}$ of the SiC sheet,
being dominated by a broad $\pi+\sigma$-electron plasmon peak at near 21 eV
and a small $\pi$-electron plasmon peak at $\sim$3 eV.
\end{abstract}


\maketitle
\section{INTRODUCTION}
Since their discovery in 1991~\cite{Iijima}, carbon nanotubes (CNTs) have attracted 
considerable interest worldwide because of their unusual properties and great 
potentials for technological
applications. CNTs can be regarded as a layer of graphene sheet rolled up in a tubular 
form~\cite{Saito}, and the structure of a CNT is specified by a chiral vector defined 
by a pair of integers ($n,m$). CNTs can be chiral or
nonchiral depending on the way they are rolled up. 
%They can be semiconducting, metallic or insulating in three types,
%i.e. armchair (n,n), zigzag (n,0), and chiral (n,m) with n$\neq$m. 
Their physical properties, in particular, optical dielectric functions, depend sensitively
on their chirality, i.e., the ($n, m$) indices (see, e.g., Ref. \onlinecite{Chu,Guo} and
references therein). Apart from CNTs, inorganic tubular materials,
such as BN~\cite{Lin,Rubio}, AlN~\cite{Mei}, GaN~\cite{Lee}, have
also been predicted and synthesized. These tubular materials also display some very interesting
properties distinctly different from their bulks. 
%The single-walled CNTs can be either semiconducting or metallic, depending on their chirality.
%The armchair CNTs has an indirect gap, whereas the zigzag type has a direct gap.
%The direct gap of the zigzag type is very important for optoelectronic devices particularly,
%as it can imply that the zigzag structure exhibit strong photoluminescence or 
%electroluminescence which never exist in its bulk structure. 

Bulk silicon carbide (SiC) crystallizes in either a cubic or a hexagonal form, and
exbihits polytypism~\cite{mun82,wyc63}. The polytypes are made of identical hexagonal layers with
different stacking sequences. These polytypes are semiconductors with a range of band gaps,
from 2.39 eV in the zincblende (3C) to 3.33 eV in the wurtzite polytype (2H)~\cite{mun82,iva92}. 
Furthermore, 3C and 6H SiC are used for high temperature, high-power and high-frequency 
devices~\cite{C,Wang,Rurali,Park} due to their unique properties~\cite{SiC}, while 6H SiC
with a band gap of 2.86 eV is a useful material for blue light-emitting diode applications~\cite{iva92}. 
Recently, SiC-NTs were also synthesized by the reaction of CNTs and SiO
at different temperature~\cite{sun02}.
This has stimulated a number of both theoretical and experimental 
investigations on the tubular form of the SiC~\cite{pei06,Pei,Menon,Xia,Mavrandonakis}.  
Based on density-functional calculations, Miyamoto and Yu~\cite{Miyamoto}
predicted the existence of graphitic and tubular forms of SiC and also proposed 
their synthesis using an extreme hole injection technique. They also reported that
the strain energies of SiC-NTs are lower than that of CNTs, and that the band
gaps of SiC-NTs can be direct or indirect, depending on the chirality. 
Using both tight-binding molecular dynamics and $\textit{ab initio}$
methods, Menon and coworkers~\cite{Menon} showed that single-walled SiC-NTs are highly stable
with a large band gap.  Zhao $\emph{et al.}$~\cite{Zhao}
also investigated theoretically the strain energy, atomic and electronic structure of SiC-NTs with 
or without hydrogenation. Gali performed an {\it ab initio} study of the effect of
nitrogen and boron impurities on the band structure of the SiC-NTs~\cite{Gali}.

Unlike CNTs, SiC-NTs are polar materials and therefore, may exhibit 
some unusual physical properties that CNTs may not have. For example,
zigzag SiC-NTs may become piezoelectric, and also show second-order
non-linear optical response.
Despite the intensive theoretical studies mentioned above,
no $\textit{ab initio}$ calculation of the dielectric response and optical 
properties of SiC-NTs has been reported, perhaps because of the heavy 
demand on computer resources. A knowledge of the optical properties of SiC-NTs
is important for their optical and electrooptical applications. 
The primary objective of this work is to
analyze the band structure and optical characteristics of all three types of the 
SiCNTs by $\textit{ab initio}$ calculations. This will help to distinguish the electronic and 
optical properties of CNTs and BN-NTs from that of SiC-NTs. Our $\textit{ab initio}$ calculations
are also needed to help understand the existing and future optical experiments. 

This paper is organized as follows. In the next Sec., the theoretical approach and 
computational method are described. In Sec. III, the band structure, density of states (DOS), 
electron energy loss spectra (EELS),
optical dielectric function of the single SiC sheet and SiC-NTs are discussed. 
Finally, conclusions are given in Sec. IV.

\section{Theory and computational method}

Our {\it ab initio} calculations for the SiC-NTs are performed using the accurate
full-potential projector augmented wave (PAW) method~\cite{blo94}, 
as implemented in the VASP package.~\cite{kre94} They are
based on density-functional theory (DFT) with the local-density approximation
(LDA). The supercell method is used such that the identical SiC tubes are 
aligned in a square array and the closest interwall distance between the 
adjacent tubes is at least 10 \AA.
We consider a number of the SiC-NTs with different diameters and chirality ($n,m$) from all 
three types, as listed in Table I.

%\subsection{\label{sec:level2}Structural optimization}

First, the ideal SiC-NTs are constructed by rolling up a graphitic hexagonal SiC sheet.
The atomic positions and lattice contants were then fully relaxed by the conjugate 
gradient (CG) method.
The theoretical equalibrium atomic positions and lattice constant were obtained when the forces
acting on all the atoms and the axial stress are less than
0.04 eV/\AA$ $ and 1.0 kBar, respectively.
In these atomic position optimizations, a uniform
grid (1$\times$1$\times$n) along the nanotube axis ($z$-axis) with the number $n$
of the $k$-points being from 30 to 50. The special $k$-point method and the Gaussian 
broadening technique were used for the Brillouin zone (BZ) integration. The length of 
theoretical translational vector $\textsl{T}$
and theoretical diameter $\textsl{D}$ are listed in Table I.
The curvature energy $E_c$ (total energy relative to that of
single SiC sheet) (also known as the strain energy) of the SiC-NTs
is listed in Table II.
Note that for the nanotubes
with a moderate diameter ($\ge 10$ \AA), the equilibrium structures are already
found to be almost the
same as that of the ideal nanotubes constructed by rolling-up a SiC sheet
with a Si-C bondlength of 1.771 \AA. This is consistent with the fact that
the calculated $E_c$ of the nanotubes with such a diameter is already smaller
than 0.05 eV/atom (Table II). Interestingly, like CNTs and BN-NTs, the calculated
$E_c$ of the SiC-NTs can be very well fitted with $E_c = \alpha /R^2$
with $\alpha$ = 2.004 eV$\cdot$\AA$^2$/atom, indicating that
the conventional elastic theory works well even down to such a small length
scale. $\alpha$ for the SiC-NTs is identical to that of the CNTs but
larger than that (1.248 eV \AA$^2$/atom) of the BN-NTs calculated previously
using the same band structure method~\cite{Lin}. However,
$\alpha$ of $\sim$1.75 eV \AA$^2$/atom reported in Ref. \onlinecite{Zhao}
is somewhat smaller than the present calculations.

\begin{table}
\caption{Theoretical structural parameters of the SiC nanotubes studied.
$\textit{D}$ is the diameter, $T$ is the length of translational vector, and
$\textit{N}$ is the number of atoms per cell.}
\begin{ruledtabular}
\begin{tabular}{cccc}
         &$D$ (\AA) &$T$ (\AA) &$N$ \\ \hline
 (3,3)   &5.05   &3.08   &12   \\
 (4,4)   &7.43   &3.08   &16    \\
 (5,5)   &8.45   &3.07   &20       \\
 (8,8)   &13.49  &3.07   &32    \\
 (12,12) &20.25  &3.07   &48   \\
 (15,15) &25.35  &3.07   &60   \\
 (3,0)   &3.36   &5.14   &12  \\
 (4,0)   &4.13   &5.22   &16  \\
 (5,0)   &5.10   &5.25   &20  \\
 (6,0)   &6.06   &5.26   &24  \\
 (8,0)   &7.89   &5.29   &32  \\
 (9,0)   &8.89   &5.3    &36  \\
 (12,0)  &11.82  &5.3    &48  \\
 (16,0)  &15.61  &5.3    &64  \\
 (20,0)  &19.85  &5.4    &80  \\
 (24,0)  &23.48  &5.3    &96  \\
 (4,2)   &5.25   &14.28   &56  \\
 (6,2)   &7.16   &19.5    &104 \\
 (8,4)   &10.5   &14.03   &112   \\
 (10,4)  &12.17  &11.06   &104
\end{tabular}
\end{ruledtabular}
\end{table}

%\subsection{\label{sec:level3}Calculations of the Band structure}

The final self-consistent electronic structure calculations were then performed 
on the theoretically determined structures of the SiC-NTs. In the
self-consistent electronic structure calculations, the number $\textit{n}$ of 
the $k$-points along the $z$-axis used ranges from 40 to 50. The density of states (DOS) was
evaluated from the final band structures by Gaussian broadening method:
\begin{equation}
N(\epsilon)=\sum_{n,\textbf{k}} w_{\textbf{k}} \delta(\epsilon -\epsilon_{n,\textbf{k}})
\end{equation}
where the Dirac delta function $\delta(x)$ is approximated by the Gaussian function,
\begin{equation}
 \delta(x) \approx \frac{1}{\sqrt{\pi}\Gamma} e^{-x^2/\Gamma^2},
\end{equation}
$w_{\textbf{k}}$ is the weight associated with $k$-point $\textbf{k}$, 
and $\epsilon_{n,\textbf{k}}$ is the $n$th energy band. 
Here the Gaussian width $\Gamma$ is set to 0.04 eV. An denser $k$-point grid
along the z-axis with $\textit{n}$ ranging from 40 to 100 
was used for the DOS calculations.

%\subsection{\label{sec:level4} Analysis of the optical properties}

The optical properties were calculated based on the independent-particle
approximation, i.e., the quasi-particle self-energy corrections, 
the local-field corrections and the excitonic effects were neglected. 
In particular, the band gap of SiC in the wurtzite (2H) structure 
from our LDA calculations is 2.15 eV, being 1.18 eV smaller than
the experimental value~\cite{mun82,iva92}. This suggests that the quasi-particle
self-energy corrections to the optical peak positions may amount to $\sim$1 eV
in the SiC systems. Furthermore, because the LDA underestimates the energy gaps, the
calculated dielectric function would be slightly too large in general. For example,
our calculated dielectric function at 0 eV for the electric field parallel and perpendicular
to the $c$-axis of 2H SiC is 8.33 and 7.90, respectively, being slightly larger
than the corresponding measured $\epsilon_{\infty}$, 6.84 and 6.51.~\cite{mun82}
Therefore, the results from the present LDA-independent-particle calculations 
should not be quantitatively compared with experiments, though they 
would certainly be useful to study the trends and characteristics of the 
optical properties of the SiC-NTs. 
Nonetheless, our previous calculations show that the dielectric functions of
graphite~\cite{Chu} and also of h-BN~\cite{Lin} calculated within the independent-particle 
picture are in reasonable agreement with experiments, perhaps because of the accidental
cancellation of the self-energy corrections by the excitonic effects. 

The imaginary part of the dielectric function $\epsilon$($\omega$) is given by the Fermi 
golden rule (see, e.g., Ref. ~\onlinecite{Chu}) (atomic units are used in this paper), i.e.,
\begin{equation}
\epsilon''_{aa}(\omega)=\frac{4\pi^{2}}{\Omega \omega^{2}} \sum_{i\in VB,j\in CB} \sum_{\textbf{k}}\ w_{\textbf{k}}
|p_{ij}^{a}|^{2}\delta(\epsilon_{\textbf{k}j}-\epsilon_{\textbf{k}i}-\omega),
\end{equation}
where $\omega$ is the photon energy and $\Omega$ is the unit cell volume. The dipole 
transition matrix elements 
$p_{ij}^{a}$=$\langle$$\textbf{k}i|$$\hat{p_{a}}$$|\textbf{k}j$$\rangle$ 
were obtained from the final self-consistent band structure with the PAW formalism.~\cite{ado01}
VB and CB denote the valence and conduction 
bands, respectively. The $|\textbf{k}n$$\rangle$ represents the
$n$th Bloch state wave function with crystal momentum $\textbf{k}$, and $\textit{a}$ represents 
the Cartesian component. We can obtain the real part of the dielectric function 
by the Kramer-Kronig transformation,
\begin{equation}
\epsilon'(\omega)=1 +\frac{2}{\pi} \textbf{P} \int_{0}^{\infty} d\omega'  \
\frac{\omega'\epsilon''(\omega')}{\omega'^{2}-\omega^{2}},
\end{equation}
where $\textbf{P}$ denotes the principal value of the integral. Given the complex 
dielectric function
$\epsilon^{'}$+i$\epsilon^{"}$, all other linear optical properties such as reflectivity,
refraction index, and absorption spectrum can be calculated. The electron energy loss spectra 
at the long wavelength limit is given by $-Im[(\epsilon^{'}+i\epsilon^{"})^{-1}]$.
The electric polarizability $\alpha(\omega)$ is 
defined by $\epsilon'(\omega)$=1+4$\pi$$\alpha(\omega)$/$\Omega$.

In the present calculations, the $\delta$-function in Eq. (3) is approximated by Gaussian 
function $[$Eq. (2)$]$ with $\Gamma=0.2$ eV. The same $k$-point grid as in the DOS
calculations is used. As noticed before~\cite{Chu,Lin}, the unit cell volume $\Omega$ 
is not well defined for nanotubes.
Therefore, as in previous calculations~\cite{Chu,Lin}, we used the effective unit cell volume 
for the SiC-NTs instead of the volume of the supercells, which is arbitrary. 
The effective unit cell for a nanotube is given by
$\Omega = \pi  [(\frac{D}{2}+\frac{d}{2})^{2}-(\frac{D}{2}-\frac{d}{2})^{2}]T = \pi DdT$, 
where $\textit{d}$ is the thickness of the SiC-NT cylinder which is set to the interlayer 
distance of h-SiC (3.51 \AA). 
%$T$ and $D$ are the length of translational vector and the diameter of the SiCNTs, respectively. 
Although the value of the effective unit cell volume used may change the magnitude of the 
dielectric functions as can be seen from Eq. (3), it would not affect the energy positions 
and shapes in the dielectric functions.

\section{RESULTS AND DISCUSSION}

\subsection{Single SiC sheet}

In order to understand the calculated properties of the SiC-NTs, 
%properticompare with the SiC-NTs and evaluate the accuracy of the present
%independent-particle approach to the optical 
%properties of the SiC structures, 
we have also calculated the self-consistent electronic band structure and 
dielectric function for an isolated 
honeycomb SiC sheet. The isolated SiC sheet is simulated by a slab-supercell approach with an 
intersheet distance of about 10 \AA. The underlying structure was determined theoretically
by using a $k$-mesh of $10\times 10\times1$, and the theoretical lattice constant is
$a = 3.069$ \AA$ $. 

The calculated band structure and density of states for the isolated SiC sheet
are displayed in Fig. 1. Clearly, the SiC sheet is a semiconductor with a direct band gap
of 2.58 eV. 
The top of valence bands and the bottom of conduction bands both occur at symmtry point 
K in the hexagonal Brillouin zone.
%In order to facilitate the discussion below, we briefly summarize the salient
%features of the present band structure of an isolated SiC sheet and h-SiC. 
Both the upper valence band and the lower conduction band are predominantly of Si and 
C $\textit{p}$-orbital character. The upper valence band 
consists of one $\pi$ band 
which arises from the $2p_{z}$ and $3p_{z}$ orbitals, extending above and below 
the SiC layer plane, and two $\sigma$
bands involving the three Si 2$s$, $2p_{x}$, $2p_{y}$ and three C 3$s$, $3p_{x}$, $3p_{y}$ 
orbitals, which form the coplanar Si-C $\sigma$ bonds. The low-lying 
conduction bands ranging from 2.4 eV to 5.6 eV are predominantly of the $p_{z}$ character,
indicating that they are the $\pi^{*}$ bands. 
These electrons play an essential role for electrical conductivity. 

The calculated dielectric function $\varepsilon$($\omega$) of 
the SiC sheet is displayed in Fig. 1. In our calculations, a dense $k$-point grid of
60$\times$60$\times$1 is
used. The optical properties of the single SiC sheet can be roughly 
divided into two spectral
region. In the low-energy range from 2 to 6 eV, the interband optical transitions involve
mainly the $\pi$ bands. At higher energies, the optical absorption peaks 
between 6 to 11 eV are associated
with the interband transitions involving $\sigma$ bands.
Strong anisotropy in the optical spectra is expected due to distinct optical selection rules,
as can be seen from Fig. 1. 
For a single SiC sheet, only $\pi$$\rightarrow$$\pi^{*}$ and $\sigma$$\rightarrow$$\sigma^{*}$ 
transitions are allowed if the electric field
$\vec{E}$ is polarized parallel to the SiC-layer plane ($E\|\hat{a}$).
In contrast, only $\pi$$\rightarrow$$ \sigma^{*}$ and $\sigma$$\rightarrow$$ \pi^{*}$ transitions 
are allowed if the $\vec{E}$ is polarized perpendicular to the SiC-layer plane ($E\bot\hat{a}$).
This explains why there is a strong absorption peak at $\sim$ 3.0 eV for ($E\|\hat{a}$)   
in the single SiC sheet [Fig. 1(c)]. For the single SiC sheet, there is no optical
absorption for $E \parallel \hat{c}$ below $\sim$5 eV (Fig. 1c).

The calculated electron energy loss spectra (EELS) of the SiC sheet are shown
in Fig. 1e. For $E \parallel \hat{a}$, two prominent peaks are found
in the energy loss function, $-Im\epsilon^{-1}$ (Fig. 1e).
A small one at $\sim$5 eV may be attributed to the
collective excitation of $\pi$ electrons partially screened by the $\sigma$ electrons.
A large broad resonance (actually, a multi-peak manifold) near 21.0 eV 
is associated with plasma oscillations involving both the
$\pi$ and $\sigma$ electrons. For $E \parallel \hat{c}$,
there is also a broad resonance in the high energy region, but at 
around 23.5 eV. However, there is 
no distinct peak at $\sim$5 eV. There are instead
many spiky thin peaks in the low energy region from 7 to 15 eV (Fig. 1e). 
%The EELS of the h-SiC for $E\|\hat{c}$ is very similar to that
%of $E\|\hat{a}$. The weak anisotropy
%in the EELS of the h-SiC may be attributed to the fact that the optical transitions
%probabilities of
%$\pi$$\rightarrow$$ \pi^{*}$, $\sigma$$\rightarrow$$ \sigma^{*}$ are almost the
%same to these of the
%$\pi$$\rightarrow$$ \sigma^{*}$, $\sigma$$\rightarrow$$ \pi^{*}$. 
%The EELS of the SiC sheet show a strong
%anisotropy. The EELS of the SiC sheet for $E\|\hat{a}$ has a broad peak from 10 to 20 eV.
%It may mean that all optical
%transitions are possible for $E\|\hat{a}$. On the other hand, the EELS of the SiC
%sheet for $E\|\hat{c}$ has several
%distinct peaks. We find that the differential of these peaks for $E\|\hat{c}$ is
%about 2 eV, it may imply that electrons
%$\pi$$\rightarrow$$ \sigma^{*}$, $\sigma$$\rightarrow$$ \pi^{*}$ are allowed if
%the electric field is polarized
%perpendicular to the sheet plane($E\|\hat{c}$).

%For $E \parallel \hat{a}$, two prominent peaks are found
%in the energy loss function, $-Im\epsilon^{-1}$ (Figs. 3e and 3f).
%A small one at $\sim$8.0 eV has been attributed to the
%collective excitation of $\pi$ electrons partially screened by the $\sigma$ electrons.
%A large broad resonance near 26.0 eV is associated with plasma oscillations
%involving both the
%$\pi$ and $\sigma$ electrons. On the other hand, for $E \parallel \hat{c}$,
%there is no distinct peak at $\sim$8.0 or $\sim$26.0 eV. There are instead
%many weak resonance in the energy range from 10.0 eV (5.0 eV for
%$h$-BN) upwards (Fig. 3e and 3f).

\subsection{\label{sec:level7}Band structure of the SiC nanotubes}

The calculated band gaps of all the SiC-NTs are listed in Table II and also displayed
as a function of the tube diameter in Fig. 2. The band structures of the selected zigzag, 
armchair and chiral SiC-NTs are shown
in Figs. 3 and 4. All the SiC-NTs, with the exceptions of
the ultrasmall (3,0) and (4,0) nanotubes, are semiconductors. In particular, all the
semiconducting zigzag SiC-NTs have a direct band gap (Table II and Fig. 3). 
The top of valence bands and the bottom of the conductions both occur at the 
center ($\Gamma$) of the 1D Brillouin zone.
This suggests 
that the zigzag SiC-NTs may find applications in optical and opto-electronic devices such
as SiC lasers.  
Interestingly, the ultrasmall diameter zigzag (3,0) and (4,0) tubes are metallic
(Fig. 3a-b and Table II). The (3,0) nanotube has a light electron pocket
at Z with an effective mass of 0.27 and two doubly degenerate
heavy hole pockets at $\Gamma$ with an effective mass of 1.0.
The (4,0) nanotube has a light electron pocket at  $\Gamma$ with an effective mass 
of 0.51 and two doubly degenerate heavy hole pockets at Z with an effective mass of 1.42.
They could be the smallest SiC metallic nanowires
and may have applications in nano-electronics and high efficient field emissions.
Nevertheless, since the LDA underestimates the band gap, this prediction of
metallic (3,0) and (4,0) zigzag SiC nanotubes should be treated with caution.
In contrast, all the armchair and chiral SiC-NTs are indirect band gap semiconductors
(see Table II and Figs. 4). For the armchair SiC-NTs, the bottom of conduction
bands generally appears at the 1D Brillouin zone boundary (Z),
while the top of valence bands occurs at somewhere between the $\Gamma$ and Z points (Fig. 4).

Fig. 2 shows that the band gap of
the small SiC-NTs increases with diameter, and approaches the band gap of the 
isolated SiC-sheet as the diameter becomes larger than $\sim$20 \AA. The reduction 
of the band gap in the small diameter SiC-NTs may be attributed to the curvature effects~\cite{Okada}.
When a SiC sheet is rolled up to form a SiC-NT, the $\pi$ and $\sigma$ orbitals are no longer
orthogonal to each other, and can hybridize. The hybridization of the $\pi$ and $\sigma$ orbitals
would modify the band structures of the SiC-NTs obtained by rolling up a SiC-sheet.
Remarkably, the reduction of the band gaps of the zigzag SiC-NTs 
is monotonic and  very large such that the ultrasmall (3,0) and (4,0) zigzag tubes become
metallic (Fig. 2). This strong diameter dependence of the band gap can perhaps
be used to engineer the band gap of the SiC-NTs by growing the tubes with a prespecified diameter.
Nonetheless, the reduction of the band gap of the armchair SiC-NTs is much smaller 
(within 0.5 eV), except the (4,4) SiC-NT (Fig. 2).
The reduction of the band gap for the chiral SiC-NTs is in between the zigzag
and armchair nanotubes. A similar trend of the diameter dependence for the SiC-NTs 
was reported before in Ref. \onlinecite{Xia}, though the sizes of the calculated band gaps
reported in Ref. \onlinecite{Xia} are generally smaller than the ones presented here.
This difference may be due to the fact that the different band structure calculation methods
were used in the previous\cite{Xia} and present calculations, namely,
the linear combination of numerical atomic-orbitals method in Ref. \onlinecite{Xia} and
the plane wave expansion method in this work. Apart from the differences
in the band gaps, the calculated band dispersions in some SiC-NTs from
Ref. \onlinecite{Xia} and this work differ noticeably too. For example, in Ref. \onlinecite{Xia},
the bottom of the conduction bands of the (5,5) SiC-NT appears near the $\Gamma$
point, while it occurs at the Z point in the present calculations. 
It is gratifying that the band dispersions of the (5,5) SiC-NT reported here
(Fig. 4) is very similar to that in Ref. \onlinecite{Miyamoto} obtained also
using a plane wave expansion method.

We note that the electrical property of the CNTs depends strongly on their chirality.~\cite{Saito}
For example, all the armchair CNTs are metallic. This is because
the purely covalent honeycomb graphene sheet is a semimetal with the conduction and valence bands
touching the K points in the hexagonal Brillouin zone.~\cite{Saito} 
In contrast, all the SiC-NTs except
the ultrasmall diameter (3,0) and (4,0) nanotubes, are semiconductors. As pointed out 
by Zhao, {\it et al.},\cite{Xia} this is due to the ionicity of SiC which results in the opening
of a band gap at the K points in the Brillouin zone of the honeycomb SiC sheet (Fig. 1).
We also note that this diameter dependence of the band gap of the SiC-NTs 
(Fig. 2) is very similar to that found for the BN-NTs.~\cite{Lin} Nonetheless, 
the BN-NTs are insulators with a much wider band gap because of the much higher 
ionicity of the BN systems.~\cite{Rubio,bla94,Lin} 
 
\begin{table}
\caption{Band gap $E_g$, band gap type, and curvature energy $E_c$ [i.e.,
total energy relative to that of the SiC sheet (-7.68 eV/atom)]
of the SiC nanotubes. The band gap of the single SiC sheet is 2.58 eV and
direct.}
\begin{ruledtabular}
\begin{tabular}{cccc}
        &$E_{g}$ (eV)    &$E_{g}$ type     & $E_c$ (eV/atom) \\ \hline
% h-SiC   &0.65          &direct           &-0.01       \\ 
%SiC-sheet&2.41          &direct           &0.00      \\
 (3,3)   &2.13          &indirect         &0.21  \\  
 (4,4)   &1.60          &indirect         &0.13   \\
 (5,5)   &2.18          &indirect         &0.09  \\
 (8,8)   &2.32          &indirect         &0.04   \\
 (12,12) &2.41          &indirect         &0.02    \\
 (15,15) &2.46          &indirect         &0.01    \\
 (3,0)   &0(metal)      &-                &0.70  \\
 (4,0)   &0(metal)      &-                &0.42   \\
 (5,0)   &0.19          &direct           &0.26   \\
 (6,0)   &0.70          &direct           &0.18   \\
 (8,0)   &1.35          &direct           &0.11   \\
 (9,0)   &1.53          &direct           &0.08    \\
 (12,0)  &1.89          &direct           &0.05    \\
 (16,0)  &2.15          &direct           &0.04   \\
 (20,0)  &2.26          &direct           &0.02    \\
 (24,0)  &2.35          &direct           &0.01    \\
 (4,2)   &0.75          &indirect         &0.39     \\
 (6,2)   &1.47          &indirect         &0.12     \\
 (8,4)   &1.92          &indirect         &0.06      \\
 (10,4)  &2.05          &indirect         &0.05      \\
\end{tabular}
\end{ruledtabular}
\end{table}

\begin{table}
\caption{Static dielectric function $\epsilon(0)$, and polarizability $\alpha(0)$ per unit 
length of the SiC-NTs. The $\epsilon_{aa}(0)$ and $\epsilon_{cc}(0)$ of the single SiC sheet
are 10.31 and 4.02, respectively.}
\begin{ruledtabular}
\begin{tabular}{ccccc}
&$\epsilon_{xx}$(0)      &$\epsilon_{zz}$(0) &$\alpha_{xx}$(0) &$\alpha_{zz}$(0) \\ \hline
% h-SiC           &16.34         &6.94                  & -               & -                 \\
%SiC-sheet        &10.88         &4.02                  & -               & -  \\
% (3,3)           &60.77         &9.82                  &264.54            &39.03  \\
 (4,4)           &7.88          &6.73                  &44.76             &37.33   \\
 (5,5)           &6.89          &9.88                  &43.61             &67.51     \\
 (8,8)           &6.91          &10.19                 &68.15             &108.64   \\
 (12,12)         &6.89          &9.97                  &104.57            &159.19   \\
 (15,15)         &6.91          &9.96                  &131.28            &199.02   \\
 (3,0)           &11.81         &8.45                  &31.81            &21.93     \\
 (4,0)           &6.47          &9.18                  &19.79            &29.59     \\
 (5,0)           &6.25          &11.46                 &23.46            &46.73     \\
 (6,0)           &6.42          &10.97                 &28.76            &52.72    \\
 (8,0)           &6.55          &10.44                 &38.86            &66.03   \\
 (9,0)           &6.66          &10.39                 &44.06            &73.21   \\
 (12,0)          &6.76          &10.58                 &59.69            &96.07     \\
 (16,0)          &6.88          &10.27                 &80.37            &126.83  \\
 (20,0)          &7.16          &10.72                 &102.13           &154.01   \\
 (24,0)          &6.89          &10.17                 &121.07           &188.67   \\
 (4,2)           &7.49         &11.34                 &27.87            &47.55     \\
 (6,2)           &6.88         &10.07                 &37.21            &57.39\\
 (8,4)           &6.87         &10.03                 &53.02            &81.49 \\
 (10,4)          &6.89         &10.04                 &62.81            &96.34 \\
\end{tabular}
\end{ruledtabular}
\end{table}

\subsection{\label{sec:level8} Optical dielectric function of the SiC nanotubes}

The calculated optical dielectric functions of some selected zigzag $[(8,0),(9,0),(12,0),(16,0)]$,
armchair $[(4,4),(5,5),(12,12),(15,15)]$, chiral $[(4,2),(8,4),(10,4)]$ SiC-NTs are 
shown in Figs. 5-7, respectively. 
%Below about 2 eV (the band-gap region), the $\epsilon^{"}$ is zero as it should, and
%$\epsilon'$ tends to a constant as the photon energy approaches zero.
The spectra can be roughly divided into two
regions, namely, the low-energy one from $\sim$2 to 6 eV and the high-energy
one from 6 to 15 eV. Below about 2 eV (the band gap region), the $\varepsilon''$
is zero and $\varepsilon'$ tends to a constant as the photon energy approaches zero.
For the electric field parallel to the tube axis ($E\parallel \hat{z}$),
the $\varepsilon''$ for all the three types of the SiC-NTs
with a moderate diameter (say, $D$ $>$ 9 \AA) in the low-energy region consists of a single
distinct peak at $\sim$3 eV plus a long shoulder of $\sim$2 eV on the
higher photon energy side (Figs. 5-7).
This is very similar to the case of BN-NTs found in Ref. ~\onlinecite{Lin}.
Nonetheless, the peak in the SiC-NTs (Figs. 5-7) is about twice as high as that 
in the BN-NTs~\cite{Lin}, because the BN-NTs have a band gap
which is in general twice wide than that of the SiC-NTs.
However, this is in strong contrast to the case of CNTs in which the distinct
features have been found especially for the semiconducting
chiral nanotubes~\cite{Chu} and these features can be used to characterize the chirality
of the grown carbon nanotubes by optical means~\cite{li01,bac02}.
Nevertheless, for small diameter SiC nanotubes, the $\varepsilon''$ spectrum can deviate markedly
from the general behavior described above (see Figs. 5-7).
For example, for the armchair (4,4) nanotube,
the pronounced peak at $\sim$3 eV in the $\varepsilon''$ spectrum mentioned above 
is absent (Fig. 6a).
For the chiral (4,2) nanotube, the pronounced peak in the $\varepsilon''$ spectrum
shifts to low energy side by about 0.6 eV (Fig. 7a).
In the high-energy region, the $\varepsilon''$ for all the types of the SiC-NTs have a
broad peak of $\sim$6 eV wide centered at $\sim$8 eV.

For the electric field perpendicular to the tube axis ($E\perp \hat{z}$),
the $\varepsilon''$ spectrum of all the SiC-NTs except
the small diameter nanotubes such as the (8,0) and (4,4), in
the low energy region also consists of a pronounced peak at around 3.5 eV
whilst in the high-energy region is, roughly, made up of a broad hump starting
from 6.0 eV (Figs. 5-7). The magnitude of the peaks is in general about
half of the magnitude of the corresponding ones for $E\parallel \hat{z}$, showing
a moderate optical anisotropy in the SiC-NTs.
In nanotubes, the electric field perpendicular to the nanotube axis
is generally strongly screened~\cite{aji94,Benedict,mar03,koz06,guo07}, and this is known as
the depolarization effect which is not taken into account in the present
calculations. The depolarization effect may substantially reduce the
magnitude of the $\varepsilon''$ spectrum
for $E\perp \hat{z}$~\cite{mar04}, and hence enhance the optical anisotropy.

Let us now compare the optical dielectric function of the SiC-NTs with that of
the single SiC sheet. It is clear from Fig. 1 and Figs. 5-7
that the $\varepsilon''$ spectrum for $E\parallel \hat{z}$ of the SiC-NTs
is very similar to that of the single SiC sheet for $E\perp \hat{c}$.
This is particularly true for the large or even moderate diameter
SiC-NTs in which the curvature effect
is small. This is perhaps not surprising because
the electric field polarization is parallel to the SiC layer in both
cases. However,
the $\varepsilon''$ spectrum for $E\perp \hat{z}$
is rather different from that of $E\perp \hat{c}$ of the single SiC sheet.
In particular, the $\varepsilon''$ spectrum for $E\perp \hat{c}$ for the photon
energy below $\sim$5 eV is zero in the single SiC sheet (Fig. 1), 
whilst in contrast, the $\varepsilon''$ spectrum for $E\perp \hat{z}$
of the SiC-NTs has a pronounced peak at $\sim$3.5 eV (Figs. 5-7).
This perhaps can be explained as follows.
When $E\perp \hat{z}$, it is clear that for some parts of the tube wall,
the electric field is nearly perpendicular to the SiC layer whilst for
the other parts of the tube wall, it is roughly parallel to the SiC layer.
Therefore, the dielectric function for $E\perp \hat{z}$ may be regarded
as a mixture of the dielectric functions for both  $E\perp \hat{c}$ and  $E\parallel \hat{c}$
of single SiC sheet.
This can be seen from inspection of Fig. 1 and Figs. 5-7, and is especially
clear for the large diameter SiC-NTs.
As a result, the calculated optical anisotropy of the large SiC-NTs
is smaller than that of the single SiC sheet.
Nevertheless, as mentioned before, the depolarization effect for $E\perp \hat{z}$ in the SiC-NTs
may considerably enhance the optical anisotropy of especially small diameter SiC-NTs.
Therefore, in optical experiments,
one may still see a rather strong optical anisotropy because
the $\varepsilon''$ spectrum for $E\perp \hat{z}$ would be substantially
reduced due to the depolarization effect.

\subsection{ Static dielectric response of the SiC nanotubes}

The optical dielectric function $\epsilon(0)$ and electric polarizability $\alpha(0)$ 
in the zero frequency limit of the SiC-NTs are listed in Table III, and also displayed in
Figs. 8 and 9. Note that the $\epsilon(0)$ [$\alpha(0)$] is the electronic contribution only
[i.e., $\epsilon_{\infty}$ ($\alpha_{\infty}$)], and hence not the full static 
dielectric response [i.e., $\epsilon_{static}$ ($\alpha_{static}$)] which also contains
the ionic contribution. In Ref. \onlinecite{guo07}, it was found that 
including the ionic contribution would not change the observed trends, though 
the magnitude of the ionic contribution may be considerable. 

Fig. 8 shows that to a first order approximation,
the static dielectric constant $\epsilon(0)$ is almost a constant 
(i.e., independent of the diameter). The average value of $\epsilon(0)$ 
is 10.38 and 6.79 for $E\|\hat{z}$ and $E\bot\hat{z}$, respectively.
The former is very close to $\epsilon_{aa}(0)$ of the single
SiC sheet, while the latter is close to the circular average of the
response of each part of the SiC-NT surface, which may be approximated by
0.5($\epsilon_{aa}(0)$+$\epsilon_{cc}(0)$) with quantities for the
single SiC sheet (see Table III).
Note that the static dielectric constants $\epsilon(0)$ of the SiC-NTs
are about two times larger than that of the BN-NTs ($\sim$4.6 for $E\|\hat{z}$
and $\sim$3.6 for $E\bot\hat{z}$)~\cite{Lin}.
Therefore, the SiC-NTs may be better dielectric materials than the BN-NTs.
It is clear from Fig. 8 that there is a pronounced anisotropy in the 
static dielectric response of the SiC-NTs.
Interestingly, there is a small but discernable chirality dependent
oscillation centered at the average value of $\epsilon(0)$ 
for both electric field polarizations. In particular, for $E\|\hat{z}$,
the zigzag SiC-NTs have a slightly larger dielectric constant than
the armchair ones. 

The electric polarizability $\alpha(0)$ per unit length 
for both $E\|\hat{z}$ and $E\bot\hat{z}$ is
proportional to the tube diameter $D$, as shown in Fig. 9. 
By fitting a linear function $\alpha(0)$=$a_{0}D$ to the calculated $\alpha(0)$-vs-$D$ curves,
we obtain a slope $a_{0,x}$ = 5.20 for $E\bot\hat{z}$, and a slope $a_{0,z}$ = 7.81 
for $E\|\hat{z}$. As pointed out before for the BN-NTs~\cite{Lin},
this linear relation between $\alpha(0)$ and $D$ arises because the SiC-NTs are
insulators in which the valence electrons are tightly bound to the ions and consequently,
every atom of the same species has nearly the same static polarizability.
Therefore, $\alpha(0)$ is proportional to the number of atoms per unit length
which in turn is proportional to $D$. Nevertheless, the SiC-NTs exhibit 
a much stronger dielectric response than the BN-NTs since the BN-NTs have 
significantly smaller slopes ($a_{0,x}$ = 2.15, $a_{0,z}$ = 3.04).~\cite{Lin}
This is because the SiC-NTs have a smaller band gap than the corresponding BN-NTs,
as mentioned before.

Both the static dielectric constant $\epsilon(0)$ and electric 
polarizability $\alpha(0)$ are clearly 
anisotropic. The anisotropy in electric polarizability can be best seen
from the difference in the slopes between $E\|\hat{z}$ and $E\bot\hat{z}$,
mentioned above. The $\alpha(0)$ for $E\|\hat{z}$ is about 50 \% larger than
that for $E\bot\hat{z}$. The actual anisotropy may be even larger
because of the calculated $\epsilon(0)$ and $\alpha(0)$ for
$E\bot\hat{z}$ would be reduced when the depolarization effect were taken
into account.\cite{Benedict,koz06,guo07} %except for (15,15) and chiral type.

\subsection{Electron energy loss spectrum}
The calculated electron energy loss spectra of the zigzag, armchair and chiral
SiC nanotubes are shown in Fig. 10. 
First of all, in the low energy region up to 15 eV, the EELS spectra for the
all selected SiC nanotubes for both $E\|\hat{z}$ and $E\bot\hat{z}$
[except (4,4) tube for $E\|\hat{z}$] (Fig. 10), look very similar to 
that of $E\|\hat{a}$ of the SiC sheet (Fig. 1e). In particular, all the
spectra have a rather pronounced $\pi$ plasmon peak at $\sim$ 5 eV and then the
spectra grow steadily with energy. In the high energy region from 15 eV upwards,
the spectra are dominated by a large broad $\pi$+$\sigma$ plasmon peak with
the exact peak positon being dependent on the chirality and also the electric
field polarization. For the zigzag nanotubes, the energy position
of the $\pi$+$\sigma$ plasmon peak is at $\sim$22 eV for both electric
field polarizations. For the chiral (4,2) and (8,4) nanotubes, respectively,
the plasmon peak occurs at $\sim$20 and $\sim$22 eV for both $E\|\hat{z}$.
For the armchair SiC nanotubes, the energy position of the $\pi$+$\sigma$ plasmon peak 
appears to oscillate around 21.5 eV (Fig. 10 c and d).

Surprisingly, there is only weak anisotropy in the EELS spectra for all the
nanotubes (Fig. 10), in contrast to the rather pronounced anisotropy found in the
dielectric functions (Figs. 5-7).
The weak anisotropy in the EELS of the nanotubes may be attributed to
the fact that for $E\perp \hat{z}$ all the $\pi \rightarrow \pi^{\ast}$,
$\sigma \rightarrow \sigma^{\ast}$, $\pi \rightarrow \sigma^{\ast}$,
and $\sigma \rightarrow \pi^{\ast}$ optical transitions are excited
whilst the single SiC sheet only $\pi \rightarrow \pi^{\ast}$
and $\sigma \rightarrow \sigma^{\ast}$ transitions are possible
for $E\parallel \hat{c}$.

\section{Summary}

An \textit{ab initio} study of the structural, electronic and optical
properties of the SiC-NTs within density functioanal
theory in the local density approximation has been performed.
In particular, the properties of the single walled zigzag 
$[$(3,0),(4,0),(5,0),(6,0),(8,0),(9,0),(12,0),(16,0),(20,0),(24,0)$]$,
armchair $[$(3,3),(4,4),(5,5),(8,8),(12,12),(15,15)$]$, 
and chiral $[(4,2),(6,2),(8,4),(10,4)]$ 
nanotubes have been calculated. For comparison, the electronic structure and optical properties 
of the single SiC sheet have also been calculated.  
We find that all the SiC nanotubes are semiconductors with exceptions of
the ultrasmall (3,0) and (4,0) zigzag tubes which may be regarded as the
thinnest conducting SiC nanowires. Interestingly, the energy band gap
of the zigzag SiC-NTs may be reduced from the full energy gap of the SiC sheet
all the way down to zero by reducing the diameter (Fig. 2), though the band gap
for all the SiC nanotubes with a diameter larger than
$\sim$20 \AA$ $ approaches that of the SiC sheet. Furthermore, all the semiconducting 
zigzag SiC-NTs have a direct band gap. All these suggest that they may 
have interesting applications in optical and optoelectronic devices.
Nonetheless, both the armchair and chiral SiC-NTs have an indirect band gap. 

The optical properties of the SiC-NTs, as for the SiC sheet, 
can be divided into two spectral regions, namely, 
the lower energy region (0$\sim$6 eV), in which the optical transitions involve mainly
the $\pi$-bands, and the higher energy region from 6 eV upwards, where interband
transitions involve mainly 
the $\sigma$-bands. 
For $E\parallel \hat{z}$, the $\epsilon''$ for all the three types of the SiC-NTs
with a moderate diameter (say, $D$ $>$ 8 \AA$ $ for the zigzag SiC-NTs, $D$ $>$ 6 \AA$ $ 
for the chiral SiC-NTs) in the low-energy region consists of a single
distinct peak at $\sim$3 eV.
However, for the small diameter SiC nanotubes such as the (4,2),(4,4) SiC-NTs,
the $\epsilon''$ spectrum does deviate markedly
from this general behavior.
In the high-energy region, the $\epsilon''$
for all the SiC-NTs exhibit a
broad peak centered near 7 eV.
For $E\perp \hat{z}$, the $\epsilon''$ spectrum of all the SiC-NTs except
the (4,4), (3,0) and (4,0) nanotubes, in
the low energy region also consists of a pronounced peak at around 3 eV
whilst in the high-energy region is roughly made up of a broad hump starting
from 6 eV. The magnitude of the peaks is in general about 
half of the magnitude of the corresponding ones for $E\parallel \hat{z}$, showing
a moderate optical anisotropy in the SiC-NTs.

Interestingly, unlike the CNTs, the calculated static dielectric
constant $\epsilon(0)$ for all the SiC nanotubes
is almost independent of diameter and chirality 
with $\epsilon(0)$ for $E\parallel \hat{z}$ being only about 30 \% larger 
than for $E\perp \hat{z}$.
This is very similar to the case of the BN-NTs except that 
the values of $\epsilon(0)$ for the SiC nanotubes are about
two times larger than that of the BN-NTs.
The calculated electron energy loss spectra of all the SiC nanotubes
studied here for both electric field
polarizations are rather similar to that of $E\perp \hat{c}$ of the SiC sheet,
being dominated by a broad $\pi+\sigma$-electron plasmon peak at $\sim$ 21 eV
and a small $\pi$-electron plasmon peak at $\sim$3 eV.

\section{Acknowledgments}

The authors gratefully acknowledge financial supports from National Science Council,
Ministry of Economic Affairs (92-EC-17-A-08-S1-0006) and
NCTS of ROC. They also thank National Center for High-performance
Computing of ROC for providing CPU time.

%Discussions with Prof. K. C. Huang at Greenwich University, U.K. and 
%Dr. C. P. Kao are also acknowledged. 

\newpage

\begin{figure}
\epsfig{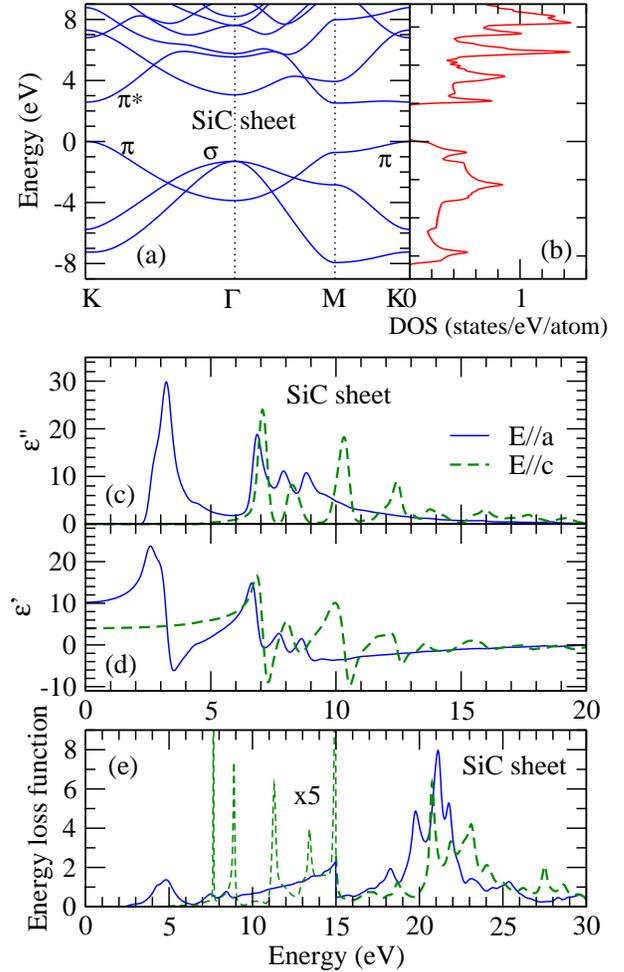}
\caption{(color online) Band structure, density of states (DOS), 
dielectric function and energy loss spectrum 
of the SiC sheet. In (a), the zero energy is at the top of 
the valence band. In (e), the spectra below 15 eV has been multiplied
by five.}
\end{figure}

\begin{figure}
\epsfig{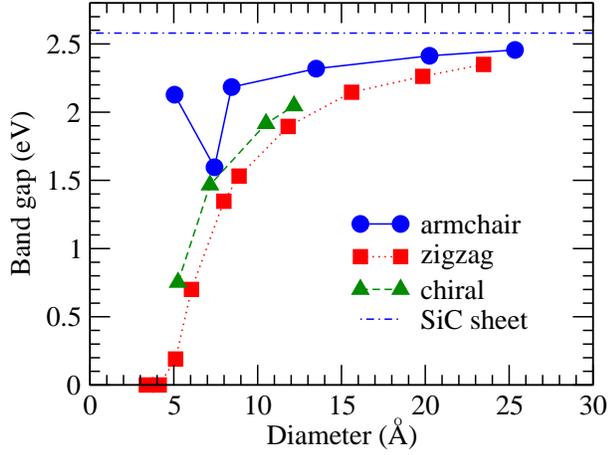}
\caption{(color online) Calculated band gaps of the SiC nanotubes vs diameter. 
For comparison, the band gap of the SiC sheet is shown as the dash-dotted line.}
\end{figure}

\begin{figure}
\epsfig{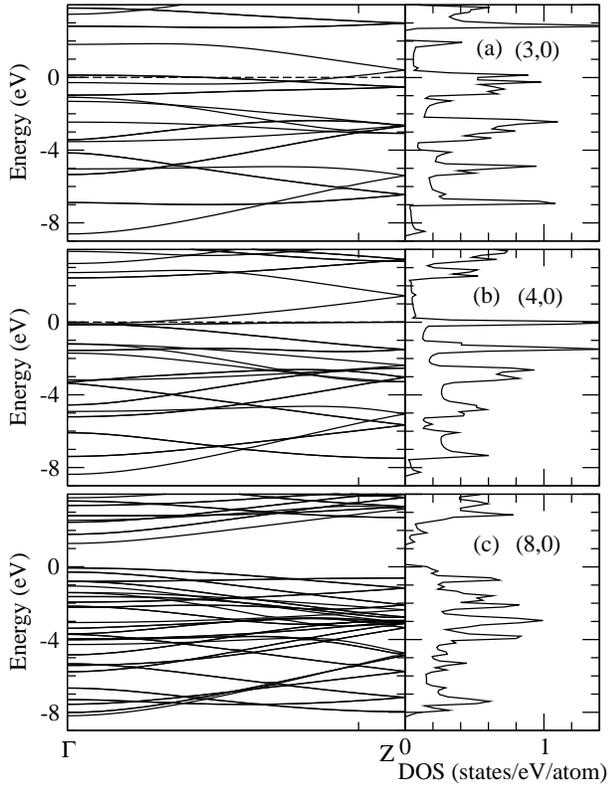}
\caption{Energy bands and density of states of the zigzag (3,0),(5,0)
(8,0) SiC nanotubes. $\Gamma$Z is the 1D Brillouin zone of length
$\pi$/$\textit{T}$ where $\textit{T}$ is the lattice constant (Table I).
In (a) and (b), the dashed line indicates the Fermi level (0 eV).
In (c) the top of the valance band is at 0 eV.}
\end{figure}

\begin{figure}
\epsfig{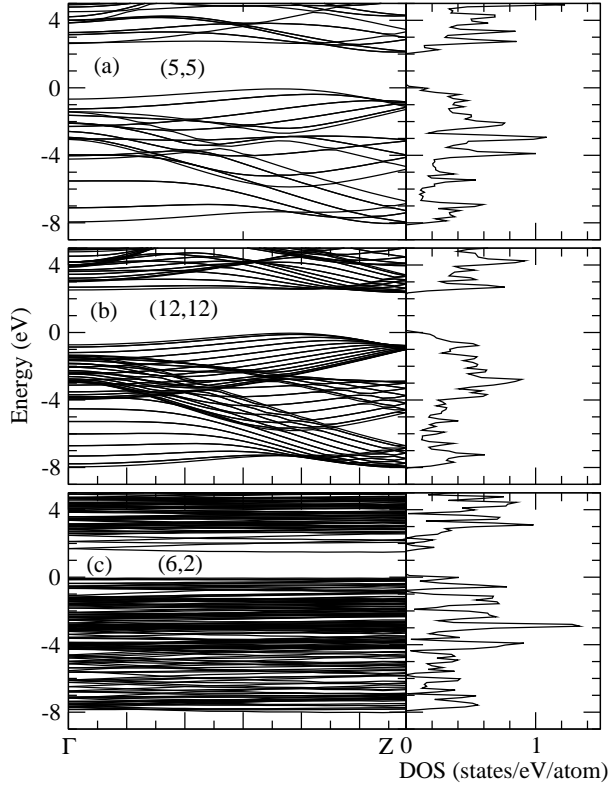}
\caption{Energy bands and density of states of the armchair (8,8), (12,12)
and chiral (6,2) SiC nanotubes. The top of the valence band is at 0 eV.
$\Gamma$Z is the 1D Brillouin zone of length
$\pi$/$\textit{T}$ where $\textit{T}$ is the lattice constant (Table I).}
\end{figure}

\begin{figure}
\epsfig{file=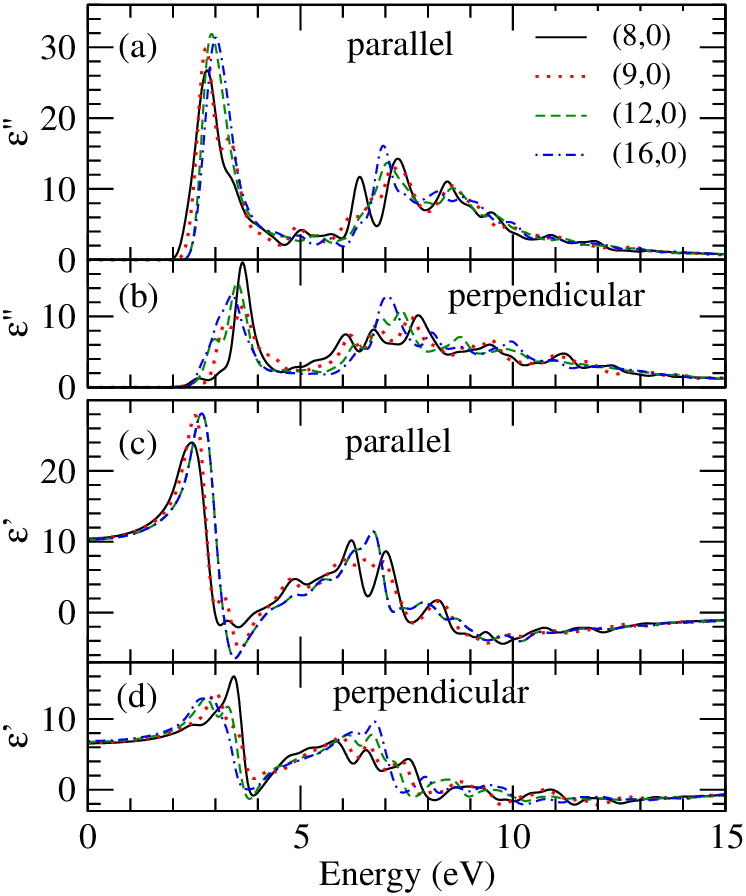,width=8cm}
\caption{(color online) Calculated dielectric functions of the zigzag SiC nanotubes.
"Parallel" and "perpendicular" represent the electric field polarized 
parallel and perpendicular to the tube axis, respectively.}
\end{figure}

\begin{figure}
\epsfig{file=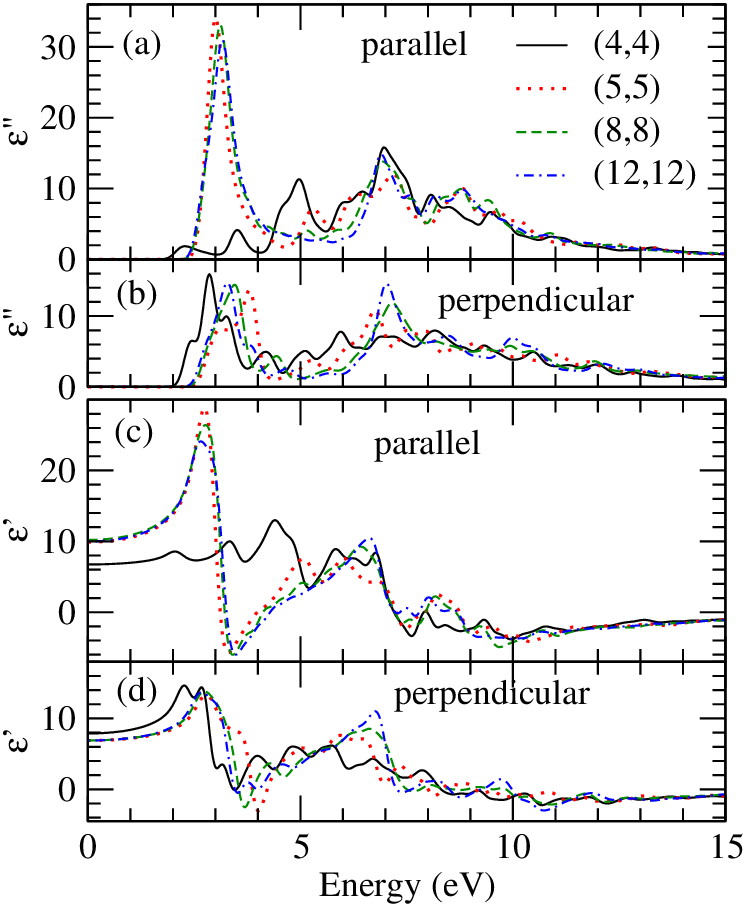,width=8cm}
\caption{(color online) Calculated dielectric functions of the armchair SiC nanotubes.
"Parallel" and "perpendicular" represent the electric field polarized
parallel and perpendicular to the tube axis, respectively.}
\end{figure}

\begin{figure}
\epsfig{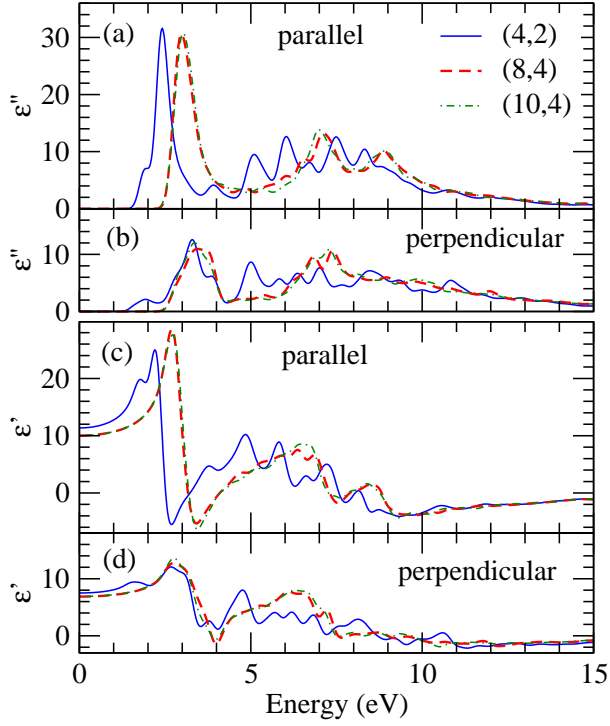}
\caption{(color online) Calculated dielectric functions of the chiral SiC nanotubes.
"Parallel" and "perpendicular" represent the electric field polarized
parallel and perpendicular to the tube axis, respectively.}
\end{figure}

\begin{figure}
\epsfig{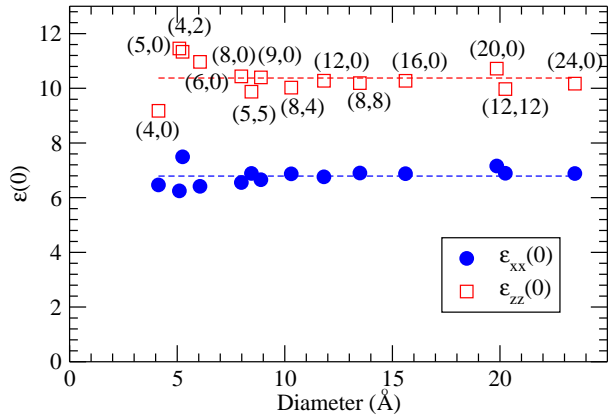}
\caption{Static dielectric constant $\epsilon(0)$ of the SiC nanotubes as a function
of diameter. The dashed line is a constant least-squares fit.}
\end{figure}

\begin{figure}
\epsfig{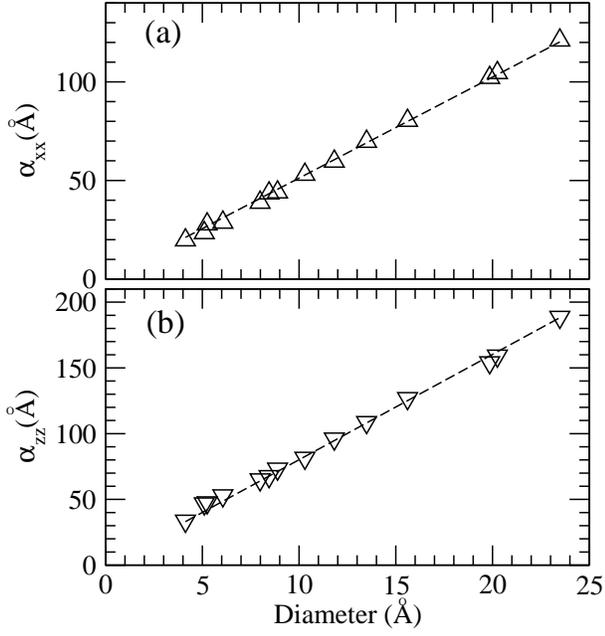}
\caption{(a) $\alpha_{xx}$(0) and (b) $\alpha_{zz}$(0) vs diameter $D$ 
for the SiC nanotubes. 
The dashed line is a linear least-squares fit.}
\end{figure}

\begin{figure}
\epsfig{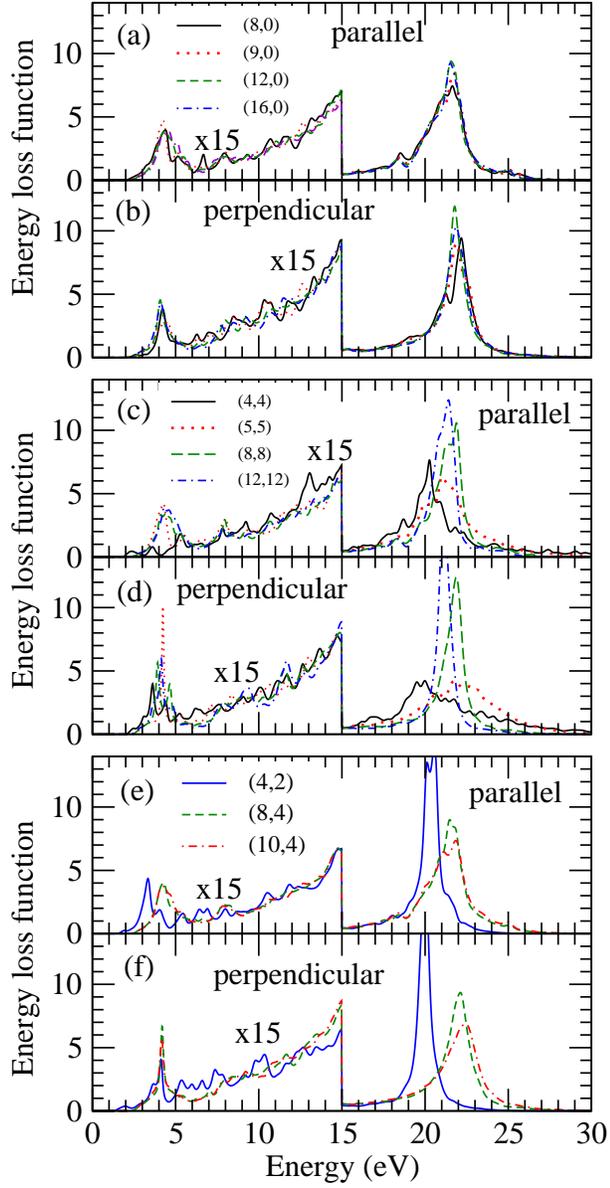}
\caption{(color online) Calculated energy loss function of the selected 
zigzag, armchair and chiral SiC nanotubes.
"Parallel" and "perpendicular" denote the electric field polarizd parallel 
and perpendicular to the tube axis, respectively.}
\end{figure}


\begin{thebibliography}{}

\bibitem{Iijima}
S. Iijima, Nature (London) {\bf 354}, 56 (1991).

\bibitem{Saito}
R. Saito, G. Dresselhaus, and M. S. Dresselhaus, Physical properties
of Carbon Nanotubes(Imperial College, London, 1998).

\bibitem{Chu}
G. Y. Guo, K. C. Chu, D. S. Wang, and C. G. Duan, Phys.Rev. B {\bf 69}, 205416 (2004).

\bibitem{Guo}
G. Y. Guo, K. C. Chu, D. S. Wang, and C. G. Duan, Comput. Matter. Sci. {\bf 30}, 269 (2004).

\bibitem{Lin}
G. Y. Guo, J. C. Lin, Phys. Rev. B {\bf 71}, 165402 (2005).

\bibitem{Rubio}
A. Rubio, J. L. Corkill, and M. L. Cohen, Phys. Rev. B {\bf 49}, 5801 (1994).

\bibitem{Mei}
M. Zhao, Y. Y. Xia, D. J. Zhang and L. M. Mei, Phys. Rev. B {\bf 68}, 235415 (2003).

\bibitem{Lee}
S. M. Lee, Y. H. Lee, Y. G. Hwang, J. Elsner, D. Porezag, and Th. Frauenheim, 
Phys. Rev. B {\bf 60}, 7788 (1999).

\bibitem{mun82} von M\"{u}nch, in {\it Landolt-B\"{o}rnstein}, edited by
O. Madelung, M. Schulz, and H. Weiss, New Series, Groups IV and III-V, Vol. 17,
Pt. A (Springer, Berlin, 1982).

\bibitem{wyc63} R. W. G. Wyckoff, {\it Crystal Structures} (Wiley, New York, 1963).

\bibitem{iva92} P. A. Ivanov and V. E. Chelnokov, Semicond. Sci. Technol. {bf 7}, 863 (1992).

\bibitem{C}
C. Persson and U. Lindefelt, J. App. Phys. {\bf 82}, 5496, (1997)

\bibitem{Wang}
R. Wang, D. Zhang, C. Liu, Chemical Phys. Lett. {\bf 411}, 333, (2005).

\bibitem{Rurali}
R. Rurali, P. Godigonon, J. Rebollo, E. Hernandez, and P. Ordejon, 
Appl. Phys. Lett. {\bf 82}, 4298 (2003).

\bibitem{Park}
C. H. Park, B. H. Cheong, K. H. Lee, and K.J. Chang, 
Phys. Rev. B {\bf 49}, 4485, (1994)

\bibitem{SiC}
$\textit{Properties of Silicon Carbide}$ edited by G. L. Harris 
(INSPEC, Institution of Electrical Engineers, London,1995)

\bibitem{sun02} X.-H. Sun, C.-P. Li, W.-K. Wong, N.-B. Wong, C.-S. Lee, S.-T. Lee,
and B.-K. Teo, J. Am. Chem. Soc. {\bf 124}, 14464 (2002).

\bibitem{pei06} L. Z. Pei, Y. H. Tang, Y. W. Chen, C. Guo, X. X. Li, Y. Yuan, and
 Y. Zhang, J. Appl. Phys. {\bf 99}, 114306 (2006).

\bibitem{Pei}
L. Z. Pei,Y. H. Tang, X. Q.Zhao, Y. W. Chen and C. Guo,
J. App. Phys. {\bf 100}, 046105 (2006).

\bibitem{Menon}
M. Menon, E. Richter, A. Mavrandonakis, G. Froudakis, and A. N. Andriotis,  
Phys. Rev. B {\bf 69}, 115322 (2004).

\bibitem{Xia}
M. Zhao, Y. Xia, F. Li, R. Q. Zhang and S. T. Lee, 
Phys.Rev. B {\bf 71}, 085312 (2005).

\bibitem{Mavrandonakis}
A. Mavrandonakis, G. E. Froudakis, M. Schnell, and Muhlhauser, 
Nano Lett. {\bf 3}, 1481 (2004).

\bibitem{Miyamoto}
Y. Miyamoto and B. D. Yu, Appl. Phys. Lett. {\bf 80}, 586 (2002).

\bibitem{Zhao}
M. Zhao, Y. Xia, R. Q. Zhang, and S. T. Lee, 
J. Chem. Phys. {\bf 122}, 214707 (2005).

\bibitem{Gali}
A. Gali, Phys. Rev. B {\bf 73}, 245415 (2006).

\bibitem{blo94} P. E. Bl\"{o}chl, Phys. Rev. B {\bf 50}, 17953 (1994); 
G. Kresse and D. Joubert, {\it ibid.} {\bf 59}, 1758 (1999).

\bibitem{kre94} G. Kresse and J. Hafner, Phys. Rev. B {\bf 47}, R558 (1993);
{\bf 49}, 14251 (1994); G. Kresse and J. Furthm\"{u}ller, 
Comput. Mater. Sci. {\bf 6}, 15 (1996).

\bibitem{ado01} B. Adolph, J. Furthm\"{u}ller, and F. Bechstedt, 
Phys. Rev. B {\bf 63}, 125108 (2001).

\bibitem{Okada}
S. Okada, S. Saito, and A. Oshiyama, Phys. Rev. B {\bf 65}, 165410 (2002).

\bibitem{bla94} X. Blase, A. Rubio, S. G. Louie, and M. L. Cohen, Europhys. Lett. {\bf 28}, 335 (1994).

\bibitem{bac02} S.M. Bachilo, M.S. Strano, C. Kittrell, R.H. Hauge, R.E.
 Smalley, and R.B. Wiesman, Science {\bf 298}, 2361 (2002).

\bibitem{li01} Z.M. Li, Z.K. Tang, H.J. Liu, N. Wang, C.T. Chan, R. Saito, S. Okada, G.D. Li, J.S. Chen, N. Nagasawa and
S. Tsuda, Phys. Rev. Lett. {\bf 87}, 127401 (2001).

\bibitem{aji94} H. Ajiki and T. Ando, Physica B {\bf 201}, 349 (1994);
 Jpn. J. Appl. Phys., Suppl. {\bf 34}, 107 (1995).

\bibitem{Benedict}
L. X. Benedict, S. G. Louie, and M. L. Cohen, Phys. Rev. B {\bf 52}, 8541 (1994).

\bibitem{mar03} A.G. Marinopoulos, L. Reining, A. Rubio, and N. Vast,
 Phys. Rev. Lett. {\bf 91}, 46402 (2003).

\bibitem{koz06} B. Kozinsky and N. Marzari, Phys. Rev. Lett. {\bf 96}, 166801 (2006).

\bibitem{guo07} G. Y. Guo, S. Ishibashi, T. Tamura and K. Terakura, 
Phys. Rev. B {\bf 75}, 245403 (2007).

%\bibitem{Deak}
%A. Gali, P. Deak, P. Ordejon, N. T. Son, E. Janzen, and W. J. Choyke, 
%Phys. Rev. B {\bf 68}, 125201 (2003).

%\bibitem{Bockstedte}
%M. Bockstedte, A. Mattausch, and O. Pankaratov,  Appl. Phys. Lett. {\bf 85}, 58 (2004).

%\bibitem{Fukumoto}
%A. Fukumoto, Phys. Rev. B {\bf 53}, 4458 (1996).

\bibitem{mar04} A. G. Marinopoulos, L. Wirtz, A. Marini, V. Olevano, A. Rubio, and L. Reining,
 Appl. Phys. A {\bf 78}, 1157 (2004).

\end{thebibliography}
\end{document}